\documentclass[epj]{webofc}
\usepackage[varg]{txfonts}   
\usepackage{hyperref}

\renewcommand*{\eqref}[1]{Eq.~(\ref{eq:#1})}

\wocname{\includegraphics[width=0.25cm,clip]{logoARENA} ARENA2018}
\wocname{ARENA2018}
\woctitle{ARENA2018}
\woctitle{\includegraphics[width=0.25cm,clip]{logoARENA} ARENA2018}
\usepackage{siunitx}
\usepackage{amsmath}
\usepackage{amssymb}
\usepackage{amsfonts}
\newcommand{\myvec}[1]{\mathbf{#1}}

\begin{document}
\title{RadioPropa --\\ A Modular Raytracer for In-Matter Radio Propagation}
%
%

\author{\firstname{Tobias} \lastname{Winchen}\inst{1}\fnsep\thanks{\email{tobias.winchen@rwth-aachen.de}; This work is funded by ERC grant 640130 and DFG grant WI 4946/1-1}}

\institute{Astrophysical Institute, Vrije Universiteit Brussel, Pleinlaan 2, 1050 Brussels, Belgium}

\abstract{%
Experiments for radio detection of UHE particles such as e.g.\ ARA/ARIANNA or
NuMoon require detailed understanding of the propagation of radio waves in the
surrounding matter. The index of refraction in e.g.\ polar ice or lunar rock may
have a complex spatial structure that makes detailed simulations of the radio
propagation necessary to design the respective experiments and analyse their
data.  Here, we present RadioPropa as a new modular ray tracing code that
solves the eikonal equation with a Runge-Kutta method in arbitrary refractivity
fields.  RadioPropa is based on the cosmic ray propagation code CRPropa, which
has been forked to allow efficient incorporation of the required data
structures for ray tracing while retaining its modular design. This allows for
the setup of versatile simulation geometries as well as the easy inclusion of
additional physical effects such as e.g.\ partial reflection on boundary layers
in the simulations. We discuss the principal design of the code as well as its
performance in example applications.
}
\maketitle

\section{Introduction}
Several experiments such as e.g.\ ARA~\cite{Allison2012},
ARIANNA~\cite{Hallgren2016} or NuMoon~\cite{Winchen2017} are currently in
preparation that use the radio emission by the Askaryan
effect~\cite{Askaryan1962} of a particle shower developing in matter to detect
high energetic cosmic particles. Design of these experiments and also future
interpretation of the collected data requires detailed simulation of the
propagation of the radio pulses from the interaction point to the receiving
antennas.
Simulations of the wave propagation in full detail with the
finite-differences-time-domain (FDTD) technique~\cite{Yee1966}
are too time consuming to be used for all tasks.  Faster solutions such as
ray-tracing or analytical estimates are thus required, that can simulate
versatile geometries as well as arbitrary models for the refractive index.

While ray-tracing can, in principle, simulate arbitrary geometries and models
for the refractive index, the underlying approximation of geometrical optics does 
not include all relevant effects. To achieve the necessary detail nevertheless,
shortcomings in the ray-tracing may be included by additional parametrizations
tuned to full FDTD simulations.  This will require a dedicated modular code for
the task and disfavours using available ray-tracing codes.

As basis for such a new code I use here the CRPropa
software~\cite{Batista2016} that has been originally designed for the
simulation of high energetic cosmic rays. The highly modular software framework
is easily extendable and already provides several necessary features such as
boundary conditions, data container, file-IO, geometry descriptions etc. - only
the required physics modules for the propagation of rays have to be created and
some trivial changes to the central data structure have to be included to trace 
rays with this software.

\section{Propagator for Gradient Media and Handling of Media Boundaries}
The path of a ray $\myvec{r}(s)$ with path parameter $s$ in a medium with index
of refraction $n(\myvec{r})$ is described by the eikonal equation
\begin{equation}
	\frac{d}{ds}\left(n(\myvec{r}) \frac{d\myvec{r}}{ds}\right) = \myvec{\nabla}n
	\label{eq:Eikonal}
\end{equation}
which is in general difficult to solve.  In the modular software presented here, boundary
effects are treated completely separate from propagation in gradient-index
media. This allows to use a locally paraxial approximation for the latter,
i.e.\ assuming that in any individual step of the algorithm the change of the
refractive index along the path $ds$ is small. Consequently eq.~\ref{eq:Eikonal}
simplifies to
\begin{equation}
	\myvec{\nabla}n \approx n(\myvec{r}) \frac{d^2\myvec{r}}{ds^2}
	\label{eq:SimplifiedEikonal}
\end{equation} which can be easily solved. For this purpose, I adapted the
Cash-Karp solver~\cite{Cash1990} already implemented in CRPropa. 

Traversing a boundary, the ray is reflected and eventually refracted according
to Snell's law.  Due to the paraxial-approximation discussed above, such
boundary conditions are treated in a separate module and consequently have to
be defined explicitly in the setup of the simulation. If the ray passes a
boundary, the ray is reflected and a secondary ray is eventually
cast from the intersection point. The amplitudes of the rays are determined
according Fresnell's equations.

\begin{figure}[tb]
	\includegraphics[width=\textwidth]{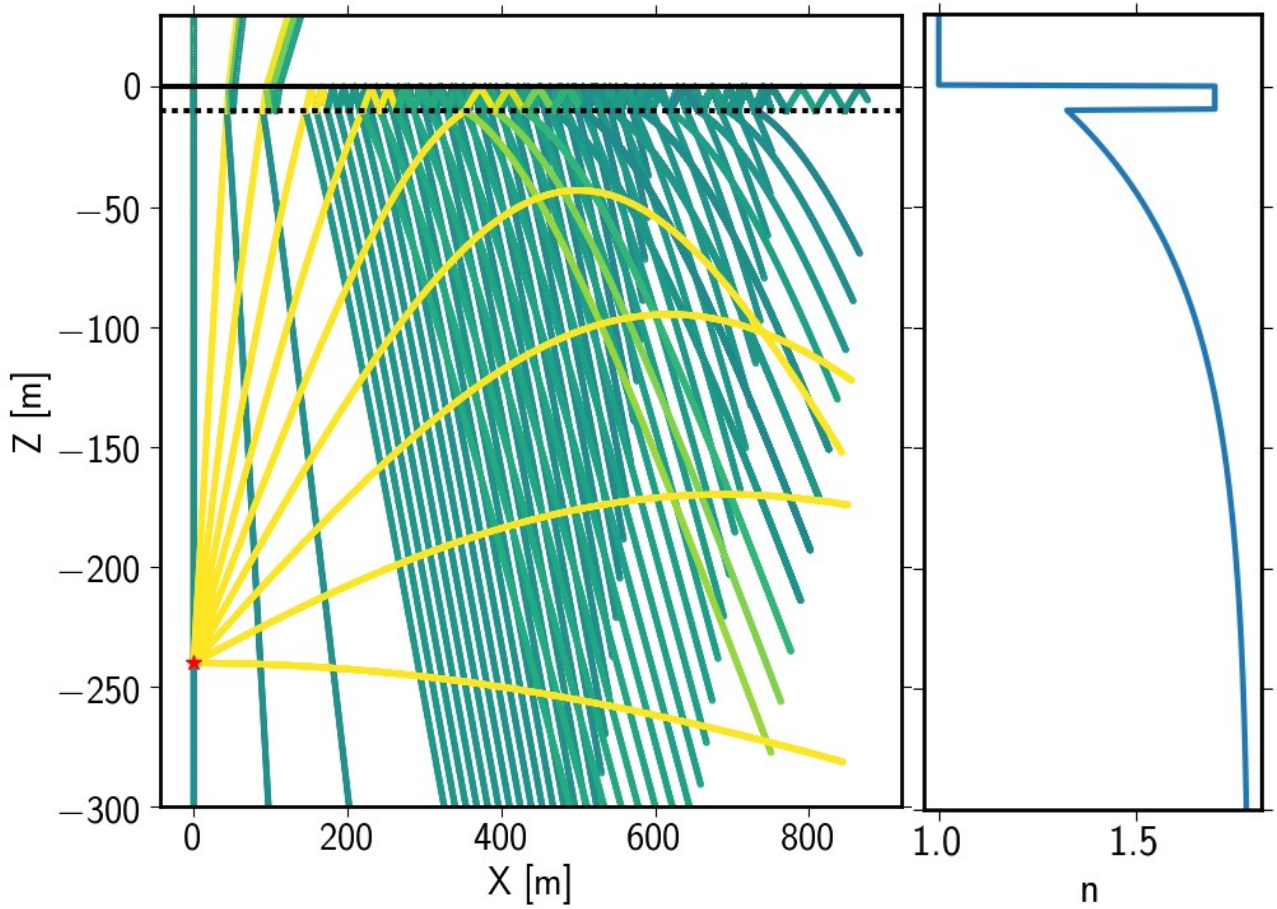}
	\caption{Example of propagated rays (left) in a two layer ice-model with
		varying index of refraction $n$ in z direction (right).  Colors denote the
		amplitude of the ray. Correction of the amplitude for length of the ray
		path is not included here to emphasize the effect of multiple reflection.
	}
	\label{fig:example}
\end{figure}
In figure~\ref{fig:example}  an example simulation for in-ice radio propagation
demonstrating these features is shown. The left part of the figures shows the
ray traces, the right part the profile of the refractive index. The model for
the refractive index is $n(z) = n_0 - \Delta n\; e^{-z/z_0}$ with $n_0 = 1.788$, $\Delta n = 0.463$ and $z_0 = \SI{71.4}{\meter}$ , below the ice-air
boundary at $z = \SI{0}{\meter}$ with an additional layer of higher refractive
index $n_1 = 1.5 $ from \SIrange{0}{10}{\meter}. Rays are emitted in several directions from
a point in \SI{240}{\meter} depth. The rays are bend in the gradient media and
reflected, respectively refracted, at the boundary surfaces as expected. Rays are
followed to a maximum propagation time of the ray \SI{5}{\micro\second} and rays with an amplitude
$A$ below the initial amplitude $A_0$ of $A/A_0 < 0.05$ are discarded.

In this example all rays are emitted in the X-Z plane for presentation clarity
only, the simulation is in three dimensions. Versatile simulation setups with
e.g.\ distributions of the refractive index from gridded data or arbitrary
equations, receiver/emitter locations, and boundary shapes as in particular
curved boundaries are also already possible with the software.

The accuracy of the simulation is determined by the paraxial-approximation of
the eikonal, the minimum step-length and tolerance of the integration. The
latter two are both free parameters of the simulation that are limited only by
the invested computing time. To investigate the accuracy I compared the result
of simulations with analytical solutions for gradient-index media with  $n(z) =
n_0 - \Delta n\; e^{-z/z_0}$ for z < 0 and $n = 1$ for $z > 0$ and parameters
as above.  I traced rays with various launch angles up to a linear distance of
1500 m from the source and compared the parameters with the analytical solution
at the observation point. The launch angles agree in this scenario to better
than \SI{0.01}{\degree}, the path length to better than \SI{0.005}{\meter} and
the travel time to better than \SI{7d-12}{\second}.

%
%
%
%
%

\section{Conclusion}
I demonstrated that the cosmic ray propagation code CRPropa is a solid base
for a radio propagation code requiring only few modifications,  that, however,
still justify a fork of the code. The resulting code `RadioPropa' is a modular
code for in matter ray tracing that handles boundary conditions and
gradient-index media with a local paraxial approximation. Due to the modular
structure, alternative propagators with higher accuracy and parametrizations for
additional effects can be easily included if needed in the future. RadioPropa is free
software published under the GPLv3 license and can be downloaded from
\url{https://github.com/TobiasWinchen/RadioPropa}.


\end{document}